%% file: conference_101719.tex
\documentclass[10pt, conference]{IEEEtran}
\IEEEoverridecommandlockouts
\usepackage{cite}
\usepackage{amsmath,amssymb,amsfonts}
\usepackage{algorithmic}
\usepackage{graphicx}
\usepackage{textcomp}
\usepackage[table, svgnames, dvipsnames]{xcolor}
\usepackage{makecell, cellspace, caption}
\usepackage{xspace}
\usepackage{comment}
\usepackage{tcolorbox}
\usepackage{array}
\usepackage{float}
\usepackage{balance}

\def\BibTeX{{\rm B\kern-.05em{\sc i\kern-.025em b}\kern-.08em
    T\kern-.1667em\lower.7ex\hbox{E}\kern-.125emX}}
    
\newcommand{\rqbox}[1]{\vspace{0.2cm}\begin{tcolorbox}[left=2pt,right=2pt,top=2pt,bottom=2pt,colback=gray!5,colframe=gray!40!black,before skip=2pt,after skip=2pt]#1\end{tcolorbox}\vspace{0.2cm}}

\newcommand{\etal}{et al.\xspace}
\newcommand{\Tool}{DiPiDi\xspace}

\begin{document}

\title{Assessing the Exposure of Software Changes:\\ The \Tool Approach*~\thanks{*This study was accepted by the MSR 2021 Registered Reports Track}}

\author{\IEEEauthorblockN{Mehran Meidani}
\IEEEauthorblockA{\textit{University of Waterloo}\\
Waterloo, Canada \\
mehran.meidani@uwaterloo.ca}
\and
\IEEEauthorblockN{Maxime Lamothe}
\IEEEauthorblockA{\textit{University of Waterloo}\\
Waterloo, Canada \\
maxime.lamothe@uwaterloo.ca}
\and
\IEEEauthorblockN{Shane McIntosh}
\IEEEauthorblockA{\textit{University of Waterloo}\\
Waterloo, Canada \\
shane.mcintosh@uwaterloo.ca}
}

\maketitle

\begin{abstract}
\input{abstract}
\end{abstract}


\section{Introduction}
\label{sec:introduction}
\input{introduction}

\section{Research Questions}
\label{sec:researchQuestions}
\input{researchQuestions}

\section{\Tool}
\label{sec:tool}
\input{tool}

\section{Research Protocol}
\label{sec:researchProtocol}
\input{researchprotocol}

\section{Threats to Validity} 
\label{sec:threats}
\input{threats}

\balance
\bibliographystyle{IEEEtran}
\bibliography{bib.bib}{}

\end{document}

%% file: abstract.tex
\textit{Context}: Changing a software application with many build-time configuration settings may introduce unexpected side-effects.
For example, a change intended to be specific to a platform (e.g., Windows) or product configuration (e.g., community editions) might impact other platforms or configurations. Moreover, a change intended to apply to a set of platforms or configurations may be unintentionally limited to a subset.
Indeed, understanding the exposure of source code changes is an important risk mitigation step in change-based development approaches.\\
\textit{Objective}: In this experiment, we seek to evaluate \Tool, a prototype implementation of our approach to assess the exposure of source code changes by statically analyzing build specifications.
We focus our evaluation on the effectiveness and efficiency of developers when assessing the exposure of source code changes.\\
\textit{Method}: We will measure the effectiveness and efficiency of developers when performing five tasks in which they must identify the deliverable(s) and conditions under which a change will propagate. We will assign participants into three groups: without explicit tool support, supported by existing impact analysis tools, and supported by \Tool.

%% file: introduction.tex
%1) Software systems that support multiple variants are complex
Complex software programs employ many compile-time configuration settings to build different software products (a.k.a., variants) from the same artifacts (i.e., source files)~\cite{tu2001build}. For example, the Linux kernel has more than 10,000 compile-time configuration settings~\cite{sincero2007linux}. These systems have multiple dependency paths to their source files from their \emph{deliverables}, i.e., software artifacts that users can interact with, such as executable files or libraries. Build systems derive default configuration settings by analyzing the execution environment or reading user override settings. Build systems use these settings to reason about whether source files (or conditionally compiled code snippets) should be included or excluded from the produced deliverables. Under some conditions, a source file may play a role in one compiled deliverable without affecting others. For example, in the Linux kernel, the source files written specifically for the ARM architecture will be excluded from the x86 version of the kernel~\cite{nadi2014linux}. In these complex systems, a change in a source-file may have unexpected side-effects on deliverables outside of the current compilation path. Software systems that support multiple variants can therefore create complex arrangements of effects and side-effects, where the deliverables exposed to a code-change can be unclear~\cite{bezemer2017empirical}.

Software engineering practices that assess source code changes, like code review, are expensive and time-consuming~\cite{cohen2010modern,Bosu2015Characteristics}. Extra time and effort must be spent by developers on activities like finding which deliverables are exposed to a change. In this paper, we define the exposure of a change as the set of deliverables affected by a change, including executables and libraries, as well as the different build-time configuration and environment settings under which the changes propagate. Changes that impact critical deliverables or configurations may require more quality assurance effort than others to mitigate their exposure risk~\cite{wen2018blimp}. 

\begin{figure}[t!]
    \includegraphics[width=\columnwidth]{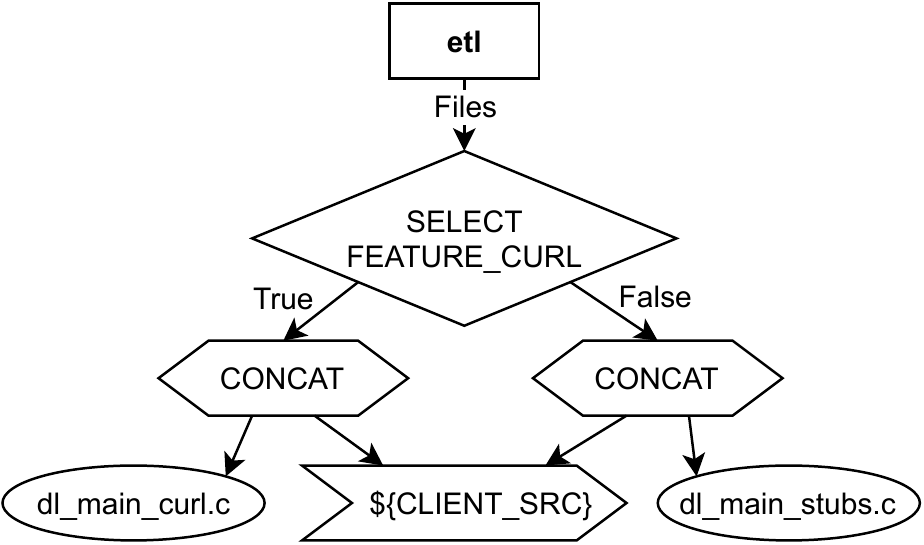}
    \caption{A real-world example of a build dependency graph}
    \label{fig:graphsample}
\end{figure}

%2) When modifying these systems such changes may be localized or broad
When modifying complex software programs, source code changes may be localized or broad.
Figure~\ref{fig:graphsample} shows an example of a dependency graph for the ET: Legacy project.\footnote{https://github.com/etlegacy/etlegacy} A change to the \texttt{dl\_main\_curl.c} file impacts the deliverable \texttt{etl} if the \texttt{FEATURE\_CURL} option is \texttt{ON}. On the other hand, changes to files represented by \texttt{\${CLIENT\_SRC}} will always impact the deliverable. A change that only impacts one variant of a system may not be as important as a change that affects all variants. Exposing the effect of a change under different configuration settings can help developers assess the impact of that change.

\begin{figure*}[t]
\centering
    \includegraphics[width=\textwidth]{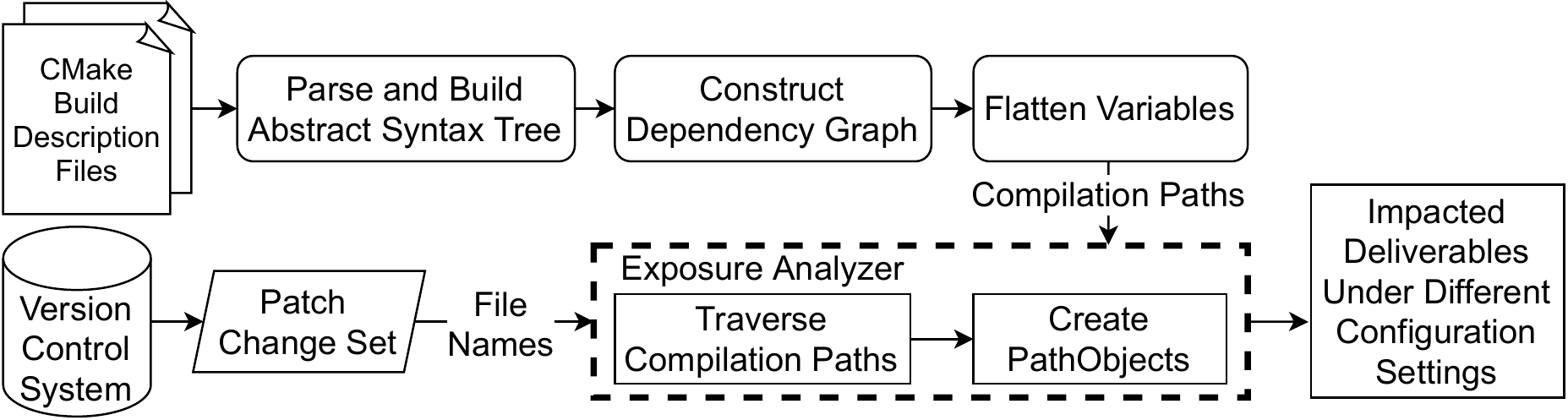}
    \caption{An overview of the \Tool approach}
    \label{fig:inspector}
\end{figure*}

Despite its importance, assessing which deliverables are impacted by a change, and the conditions under which they are impacted, is not well supported by current software tools~\cite{hassan2018hirebuild}. Change Impact Analysis (CIA) is one way to determine the consequences of a change on a software application \cite{arnold1993impact}. Many CIA techniques have been proposed~\cite{li2013survey, ahsan2010impact,gethers2010using,tamrawi2012build,adams2007design,Gyori2017Refining}. However, to the best of our knowledge, none of them consider environment or build-time configuration settings. While build impact analysis has been shown to be effective~\cite{wen2018blimp, adams2007design}, current techniques rely on a dynamic analysis of build execution, which cannot expose the impact of a change on different environmental and configuration settings.

%Ahsan \etal propose a CIA technique using information retrieval and machine learning on software change request (SCR)~\cite{ahsan2010impact}. Gethers \etal introduced a new coupling metric and use that for impact analysis~\cite{gethers2010using}. Tamrawi \etal developed \texttt{SYMake}~\cite{tamrawi2012build}, a tool that help developers to find smells and refactor build files by building a symbolic dependency graph from the GNU make. Wen \etal, shows the build impact analysis is effective~\cite{wen2018blimp}, but they use \texttt{MAKAO} tool~\cite{adams2007design} which uses build output information and does not consider different configuration settings. 

%We propose ...
Therefore, we propose \Tool, an approach to assess the exposure of changes to the source code of systems that are built using CMake. One of the key roles of the build system is finding and selecting files based on build scripts, build-time configurations, and environmental variables~\cite{zhou2014Build,Seo2014Programmers,AlKofahi2012Detecting}. By statically analyzing the build scripts and constructing the \emph{Build Dependency Graph} (BDG), we can assess the exposure of a change on all software variants.

%this paper presents...
This paper presents our plan to experimentally evaluate the effect of \Tool on the effectiveness and efficiency of determining the exposure of source code changes.
% In this experiment, we investigate whether revealing the exposure of a change under different configurations helps developers identify impacted deliverables more effectively and efficiently. 
To that end, we form three participant groups -- those with no tool assistance, those with the assistance of a commercial CIA tool, and those with the assistance of \Tool -- and compare their efficiency and effectiveness on prescribed tasks. The participants are asked to identify the impacted deliverables and variants for given source code changes while we monitor their performance. A tool that could significantly improve effectiveness and efficiency for these tasks could be useful in many applications both for researchers who design experiments based on source code change (e.g., mutation testing)~\cite{rovegaard2008empirical} and practitioners in the allocation of quality assurance resources.

%% file: researchQuestions.tex
In this study, we aim to determine whether a static analysis of build systems can improve the effectiveness and efficiency of software developers striving to assess the exposure of a source code change. A source code change, or patch, that impacts an application under a specific and rare configuration would likely not merit as much developer attention as a source code change that always impacts the application. A change that impacts more deliverables and/or configurations (high-exposure) has a broader ``surface area" and a greater potential to impact users, should a defect be introduced, than a change with low-exposure. 
%Therefore, we believe that knowing which deliverables are affected by a source code change or a patch can allow developers to make more informed decisions when making source code changes. We consider patches as a set of changes in source files that fix a bug or add a new feature.
Despite the importance of understanding exposure, it is difficult to assess without tool support.
To this end, we propose \Tool to improve awareness of the exposure of changes. We hypothesise that \Tool will allow developers to more efficiently and effectively determine the exposure of source code changes.
%which deliverables are impacted by source code changes more rapidly and more efficiently.

To test our hypothesis, we  pose the following research questions (RQs):
\rqbox{\textbf{RQ1}: Does \Tool help developers assess the exposure of source code changes more effectively?}

%Alternative for RQ1: ``Can our approach help developers identify the deliverables affected by exposure of a change more effectively?"

We address RQ1 by testing the following hypotheses:\\
\textbf{H\textsubscript{1.1}}: \Tool significantly increases the effectiveness of developers in assessing the exposure of a patch.\\
\textbf{H0\textsubscript{1.1}}: \Tool \textbf{does not} significantly increase the effectiveness of developers in assessing the exposure of a patch.\\

Additionally, 
% while finding all of the deliverables impacted by a change is important, it also is time-consuming because it requires project-wide knowledge, the relation between the files, and the build script. Developers attempting this task should therefore find the usage of the modified source code throughout the project and follow build-time configurations to identify the impact of the change. Some of these configurations may be related to the environment of the user, like the operating system. So a change may have a side-effect on one machine without appearing on others. On the other hand, build-scripts may use wildcard addressing, like \textit{*.cpp}, for the source files, making it challenging to follow a complete compilation path from a deliverable to the changed source file. Therefore, we want to determine if exposing the effect of a change on an application can improve the efficiency of the developers attempting to determine the affected deliverables. Thus,
we ask:

\rqbox{\textbf{RQ2}: Does \Tool help developers to assess the exposure of source code changes more efficiently?}
We formalize RQ2 in the following hypotheses:\\
\textbf{H\textsubscript{2.1}}: \Tool significantly increases the efficiency of developers in assessing the exposure of a patch.\\
\textbf{H0\textsubscript{2.2}}: \Tool \textbf{does not} significantly increase the efficiency of developers in assessing the exposure of a patch.

\begin{comment}
Alternative for RQ2: ``Can our approach help developers identify the exposure of a change more efficiently?"
\end{comment}

%% file: tool.tex
Our proposed solution to raise developer awareness of the exposure of changes, \Tool, works on projects that use the CMake build system. CMake is a cross-platform build system that builds deliverables from artifacts, like source files \cite{cmakewebsite}. CMake has two distinct phases. First, it generates platform-based low-level build specifications (e.g., Makefiles, Visual Studio {\#.sln} files, or Ninja files~\cite{martin2010mastering}). Then, CMake invokes the low-level build tool to build the project. 

An overview of the approach used by \Tool can be found in Figure~\ref{fig:inspector}. \Tool first parses the CMake specifications starting with the \texttt{CMakeLists.txt} file in the project root directory (i.e., the entry point for the CMake build system). We use ANTLR~\cite{parr1995antlr} to parse and build the \textit{Abstract Syntax Tree (AST)} from the CMake file. At this level, we may need to include and parse other CMake files as instructed in the \texttt{CMakeLists.txt} file.

Next, we traverse the AST to create the \textit{Build Dependency Graph}, which represents the relationship between the deliverables, source files, and the conditions in each path. Using the graph, we resolve variables to their values under different build-time configuration settings (i.e., flatten the variables). By flattening the variables, we obtain all of the possible values for each variable for all configuration settings. This information is then saved and can be accessed through an API when attempting to determine the exposure of a source code change.

Given a list of changed file names, the flattened variables can be used to traverse dependency paths to create a list of exposed \texttt{PathObjects}, the output of \Tool. A \texttt{PathObject} contains all the possible dependency paths from deliverables to the changed source files. Often in large software applications, there are build-time configuration and environmental settings that help the build system to reason about different variants of the system~\cite{liebig2010analysis, hochstein2011cost}. These settings create different dependency paths from the deliverable to the source files. An example of the output of \Tool is shown in Figure~\ref{fig:tooloutput}. This output can then be used by developers to identify which deliverables and variants are exposed by source-code changes. \Tool will be available online on our public GitHub repository.\footnote{https://github.com/software-rebels}

%% file: researchprotocol.tex
To test our hypotheses, we will conduct randomized controlled experiments with three groups. Study participants will be asked to perform a set of prescribed tasks with their usual development setup without additional help (control group), with a baseline change impact analysis tool (positive control group), and with \Tool (treatment group). We measure the effectiveness of our tool by comparing the responses of the participants with an established ground truth. We will measure the efficiency of our participants by comparing the duration of each task across the groups.

\subsection{Variables}
\label{subsec:variables}
\input{variables_table}

Table~\ref{tbl:variables} provides an overview of the study variables, which we describe below.

% \begin{figure}
% \begin{minted}
% [
% xleftmargin=1em,
% frame=lines,
% fontsize=\scriptsize,
% linenos
% ]
% {c}
% #ifdef USE_MYMATH
% #include "mysqrt.h"
% #endif
% ...
% #ifdef USE_MYMATH
%   return detail::mysqrt(x);
% #else
%   return std::sqrt(x);
% #endif

% \end{minted}
%     \caption{An example of a C source file that uses \#ifdef directive for conditional compilation}
%     \label{fig:csourcesample}
% \end{figure}

\subsubsection{Independent Variable}
In our study design, the tool support provided to the participants varies (\textit{No Tool}, \textit{With Existing Tool}, and \textit{With \Tool}). All tooling level groups will have access to the same information and interface. The only difference in access being the additional output of the Existing Tool/\Tool for the relevant groups. More specifically, each group is defined as follows:\\
\textbf{No Tool}. This group has access to the code change and other files in the project, including the build specifications. They can use their preferred development environment to perform the tasks. This group is a control group and represents the current practices used by software developers attempting to determine which deliverables are affected by a source code change.
\\
\textbf{Existing Tool}. This group has access to the same environment as the \textit{No Tool} group, as well as a state-of-the-art change impact analysis tool~\cite{kirchner2015frama}. This group is a positive control group and represents the current approaches used by software engineering research to aid software developers attempting to determine which deliverables are affected by a source code change.
\\
\textbf{\Tool}: This group--the treatment group--will have access to same environment as the \textit{No Tool} group, as well as the output of \Tool. For each changed file, the tool will print a \texttt{PathObject}. Our tool will print the impacted deliverables at the file level. Although the file granularity may overestimate the true impact of a change, it is the granularity at which the build system operates.

% However, the impacted deliverables should be identified at the code level in most cases. Thus, our tool will provide file-level hints for the participants to find those deliverables.

%\begin{figure}
%\begin{minted}
%[
%xleftmargin=1em,
%frame=lines,
%fontsize=\scriptsize,
%linenos
%]
%{json}
%{
%"dl_main_curl.c": {
%        "FEATURE_CURL": ["etl"]
%    },
%"dl_main_stubs.c":{
%        "NOT FEATURE_CURL": ["etl"]
%    },
%"common.c": {
%        "": ["etl"]
%    }
%}
%
%\end{minted}
%    \caption{An example of output of the tool based on the given graph in Figure~\ref{fig:graphsample}}
%    \label{fig:tooloutput}
%\end{figure}

\begin{figure}[t]

    \includegraphics[width=.6\columnwidth]{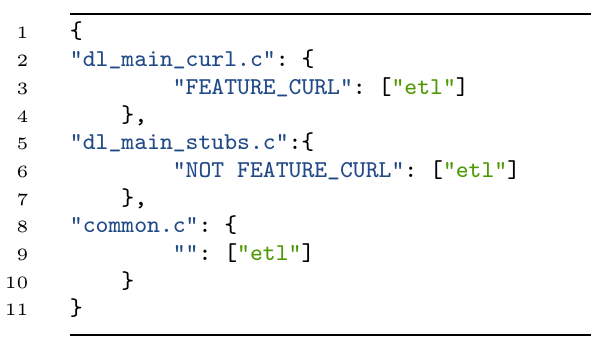}
    \caption{An example of output of the tool based on the given graph in Figure~\ref{fig:graphsample}}
    \label{fig:tooloutput}
\end{figure}

\subsubsection{Dependent Variables} Our dependent variables are outlined in Table~\ref{tbl:variables}. We discuss our reasoning for these variables below.\\
\textbf{Exposure analysis effectiveness}: The score from each task indicates how close the answers of the participants are to the ground truth. We could alternatively determine if a participant provides fully correct answers for each task and consider the ratio of correct answers to total tasks. However, we believe that our approach, which indicates how close participants are to fully correct answers, allows us to obtain a finer grained insight into how participants complete their tasks. Thus, we consider our task scores (i.e., \emph{Number of correctly identified deliverables} \& \emph{Relative rate of correctly identified deliverables}) to be good proxies for exposure analysis effectiveness.\\
\textbf{Exposure analysis efficiency}: We define exposure analysis efficiency as the scores the participant get for tasks for a given time (in minutes) spent for the tasks. As a result, getting higher score in a shorter time will result in a higher efficiency. This way, we consider both the fully correct answers and the partial ones, especially in the rank based tasks.
\begin{comment}
\textbf{Number of identified deliverables}: In order to determine whether finding the impacted variables at file-level is actually a time consuming task, we create three groups~\textit{(approaches)} to find these deliverables. We expect significant differences between the treatment and the control group.\\
\textbf{Rank changes}: Given source code patches and a build configuration, the participants should rank the patches based on the number of impacted deliverables. The control group should find the deliverables like they normally do without any metadata. Two other groups will do the same task with access to the output of a tool (either existing or proposed).\\
\textbf{Patch identification}: Given source code patches, a build configuration, and an impacted deliverable, we ask participants to list the patches that affect the deliverable. The order of the patches is not important.\\
\textbf{Rank changes (variants)}: As discussed previously, build systems take software source code and user configuration as inputs and generate different variants of a same source code. These variants may differ in features \cite{liebig2010analysis} or be the same but support different platforms \cite{hochstein2011cost}. Given few patches, participants rank them based on the number of variants that will be affected by each patch.\\
\textbf{Patch identification}: Participants should list patches that affect a specific variant of the software.
\end{comment}

\subsubsection{Confounding Variables} Because different code changes might affect the results of our participants, we control the code changes made available to them. We present patches from three different projects to ensure our results are not biased towards any single project. We also control build-time configuration settings to evaluate \emph{Tooling level} with multiple build configurations without introducing confounding factors. We gather some demographic information like the \textit{Development experience} in order to control their correlation with the dependent variables. We also use these variables to inform our data preprocessing (e.g., get some context to determine why a participant might not have finished a task) and for further analysis. We use this data to augment the statistical analysis and make decisions about whether a participant is suitable for a task.

\subsection{Materials}
\label{subsec:material}
In this section, we describe the materials that we use in this study.
\subsubsection{\Tool} We developed \Tool to reveal the exposure of a change in a structured manner. In a nutshell, \Tool processes build specifications statically to produce a \textit{Build Dependency Graph (BDG)}, which we traverse to assess exposure. Before conducting the experiments, we will run the \Tool BDG generation step on the projects that will be presented to our participants and save the output. Participants in the \Tool tooling level of the experiment will use \Tool's querying features to perform the assigned tasks.

\subsubsection{Existing tool} To assess whether the improvements in the \Tool tooling level (treatment) group are related to the approach implemented by our tool, we select a recent and available impact analysis tool to employ in the \textit{Existing tool} (positive control) group. Unfortunately, most of the proposed impact analysis tools are prototypes~\cite{li2013survey}. Additionally, due to our project selection and since our implementation of the \Tool approach supports CMake build specifications, the impact analysis tool must support the C++ programming language. We have selected Frama-C; a tool proposed by Kirchner \etal~\cite{kirchner2015frama}. Frama-C is an industrial grade static analysis tool, which can perform impact analysis on C and C++ projects. Moreover, Frama-C is open-source and can therefore be customized if needed.
\subsubsection{Studied projects} To allow for realistic evaluations, all of the patches that form the basis for the tasks in this experiment are sampled from real-world projects. Since our tool currently supports the CMake build system, we limit our selection to large and successful projects that use CMake. We choose to focus on large systems (with complex dependency graphs), since they stand to benefit more from an exposure analysis than small systems (with simple build dependency graphs) do.
%For the purpose of this study, we need to choose projects that use CMake build system and have a few build-time configuration. Often, software applications that work on different platforms and architectures have enough build-time configurations. Thus, 
We select three projects from the Qt and KDE open-source communities as the scope from which to sample tasks to conduct our study. KDE is a collection of projects comprising an open-source desktop environment. Qt is an open-source toolkit for creating Graphical User Interfaces (GUI). Both of these communities use C/C++ as their programming languages and CMake as their build technology.

To select our projects, we first start with all of the projects available on the GitHub organization pages for KDE and Qt and filter out projects that do not use CMake. We then filter out projects that have not had commits in the last 6 months to guarantee that we are looking at active projects. Finally we select the top three projects by number of forks, a proxy that allows us to gauge developer interest. With these criteria in mind, we select Kdenlive, Qt Base, and Krita.

For each studied project, we plan to iterate over patches in reverse chronological order, selecting patches that impact a different number of deliverables under different configuration settings until three patches have been selected (nine patches in total).
% For each studied project, we start from the latest change and go to backwards in time to select three patches that impact a different number of deliverables under different configuration settings.
To identify the impacted deliverables, we manually inspect the source files and find the deliverables that are impacted by the changed code. We use this as our ground truth. While \Tool reports changes at file level, in this study we ask participants to report impacted deliverables at the code level, a sub-set of reported deliverables by the tool.

\subsubsection{Experiment UI}
\label{subsec:experimentui}
To conduct our experiment with a diverse range of participants and allow our participants to rely on their own development environments, we develop a Web based application with which our participants will interact. The application will retain a log of answers and the duration of each task. The logic behind the experiment UI will randomly assign each participant to a \textit{Tooling level} group and randomly assign tasks to the participants, all the while logging which project and tasks are assigned to whom. Participant information will only be made available to the researchers after all results have been scored to reduce experimenter bias~\cite{rosenthal1976experimenter}.

\subsection{Tasks}
\label{subsec:tasks}
We ask our participants to complete five tasks, one Type A task, two Type B tasks, and two Type C tasks. After a participant initiates our experiment through our experiment UI, they are randomly assigned to a~\textit{Tooling level} and the tasks are randomly ordered and logged. The order of the tasks is randomized to account for learning effects that could occur if developers improve by learning from previous tasks. Furthermore, we construct each task using three different open-source projects, and randomly assign each task to each participant. Therefore, participants cannot share answers with each other and tasks are less biased towards a specific project or task. Participants must obtain the data and files required to complete each task through our experiment UI, and must also provide their answers through it.

Our tasks are constructed to answer both RQ1 and RQ2. The results obtained for each task can be used to answer our first research question (i.e., RQ1), while the duration of the tasks can be compared for each group to answer RQ2. The three task types are as follows. %Next we briefly talk about each tasks:
% ~\maxNote{I think we should give more 'protocol' details for each task here. More info on things like: The 'data' given to participants; The type of answer (is it just the file name or the complete path of the deliverable that we expect?); For the rank, we need to say that we do not provide patches that can allow for two equal rankings, etc. }
% ~\maxNote{We should also say which tasks target which variables.}

\textbf{Task Type A}: The purpose of this task is to compare the exposure assessment effectiveness and efficiency of the participants in different~\textit{Tooling levels}. The participant is provided with the names of changed files and a set of build specifications. The participant is then asked to list impacted deliverables (without having the source code). The experiment UI provides a text input field for the participant to identify those deliverables. 

\textbf{Task Type B}: The purpose of these tasks is to determine the effect of presenting exposure reports on the effectiveness and efficiency of developers assessing the relative exposure of patches. The participant is assigned three patches and a set of build specifications. We ask the participant to rank the patches listed in the experiment UI based on (a) the number of impacted deliverables; and (b) the number of impacted application variants (e.g., number of affected OS). We ensure that the patches do not affect the same number of deliverables and application variants. Furthermore, the patches are sampled from a different project than the ones studied for other tasks.

% \textbf{Task C}: In this task, using the same project as Task B, we ask the participant to rank the patches based on the number of affected application variants. Those variants can be a feature or a specific version of a specific operating system. These variants occur because of the different build-time configurations. Thus, the purpose of this task is to investigate the effect of exposing a source code change under different build-time configurations. Similarly to Task B, the application expects the participant to input a correctly ordered list of patches.

\textbf{Task Type C}: The purpose of these tasks is to determine the impact of \Tool when participants are particularly interested in the exposure in a given setting. Participants are presented with three patches and asked to identify those that (a) affect a specified set of deliverables; (b) affect a specific variant of the software; and (c) identify the configuration settings under which the changes will propagate. For this task type, we use a different project than for tasks of types A and B. 

\subsection{Participants}
Since our tasks are centered around specific software engineering practices, our participants should have the programming experience necessary to allow them to find the deliverables impacted by a source code change. We therefore seek to populate our pool of participants with software developers, or individuals with programming experience. 

We calculated the required size of our pool of participants using the standard settings for uncovering a medium effect size (0.25) when applying a one-way ANOVA (i.e., $\alpha=0.05$, $\beta=0.8$, three levels)~\cite{cohen1992statistical}. The results require us to recruit 159 participants. Since recruiting such a large pool of participants is unlikely, we relax our effect size target to large effect size (0.40), giving a more achievable pool of 66 participants.

We will recruit our participants for our study from the development teams of our industrial partners, which include large multinationals like Huawei and Dell EMC, as well as start-ups like YourBase. We strive to recruit at least 50 professional software developers from these organizations.

Since we intend to track project-specific experience and sample from all three types mentioned in Table~\ref{tbl:variables}, We will seek to recruit experienced software developers (i.e., contributors) in the KDE and Qt projects through personal and professional contacts, e.g., projects' mailing lists and on social media platforms like LinkedIn and Twitter. We expect to recruit at least ten software developers from these sources. Finally, we will also post open calls for participants in various schools of computer science and software engineering. We expect to recruit at least ten more participants through these sources.\\
Pilot: We will run a pilot round with five participants to test our experiment platform and the whole experiment.

\subsection{Execution Plan}
We will provide our participants with access to our web-based application in batches of ten. This staged approach will allow us to fix any potential problems without invalidating too large of a subset of our participant data. The application will have the following procedure for each participant:

\subsubsection{Welcome Page} We first provide our participants with an outline of the tasks and an estimate of the time required to complete the tasks. In addition, we will request the requisite consent of participants to participate in the experiment. The participants are asked to refrain from sharing task information with other participants. For ethical compliance reasons, participants are also informed that they may stop the experiment at any time for any reason.

\subsubsection{Onboarding}
After obtaining consent from the participants, we provide an explanation of the tasks to be completed during the experiment. Based on the tooling level assigned to the participant, we provide the steps required to setup the environment and the tool (if applicable). Additionally, we include a warm-up task, so that the participant becomes familiar with the study subject prior to taking on the evaluated task.  We inform participants that they may use their preferred development tools (e.g., CLI tools, IDE). Participants are also informed that each task is timed, that their responses will remain anonymous unless they explicitly request otherwise, and they may skip individual tasks.

\subsubsection{Tasks} We present our participants with the tasks outlined in Section~\ref{subsec:tasks} in a random order. For each task, our application will provide a hyperlink to download the source code. A timer will begin as soon as the task page is loaded. We also log checkpoints throughout the experiment. Before showing the description of the actual task, we provide the download link and the necessary steps to prepare the environment. The participants must click on a ready button to initiate the experiment. We also log the moments that the participants begin to enter their responses. The page will describe the task, and show the configuration settings that the participant should consider. 
We present the results of the tools in the experiment UI for participants in the `Existing Tool' and `\Tool' \textit{tooling level}, in a form that emulates supplementary code review information that would be available if the tools were part of a CI/CD pipeline.
% The experiment UI will also show the output of the appropriate tool for participants in the `Existing Tool' and `\Tool' groups.
The application will provide input spaces for the participant to enter their responses. The application will log the time that the participant spent on each task. The participant may click a pause button to pause the timer if a distraction of any kind interrupts their focus. A skip button allows the participant to move on if they feel that they cannot complete a task.

\subsubsection{Questionnaire} Prior to the start of the experiment, the participants are asked demographics questions about their background and programming experience. After a participant completes their five tasks, we will follow up with a questionnaire, which collects tool usage questions about the CLI tools, IDEs, and/or other tools that used to complete the tasks. Additionally, we are interested to know whether the participants find the provided tool useful and usable. We also ask participants to comment on any problems that they may have encountered during the experiment. Finally, we will thank the participants and invite them to provide other feedback if they desire. This questionnaire will take fewer than five minutes to complete.

\subsection{Analysis Plan}
We expect to see differences in effect size between the \Tool tooling level group and other groups. Based on our sample size, the differences should be large to count the effect. In this section, we describe the analysis plan we intend to use in this study.
\subsubsection{Data Cleaning} We assign each participant five tasks to complete. However, it is possible for a participant to exit the application before completing all of their assigned tasks. Since the experiment UI accepts input from participants in any text format, we will manually check that answers are sane before analyzing them. Next, we will review the participant's questionnaire submission and feedback for mentions of problems that may (partially) invalidate their submission, removing their invalid answers when appropriate. Additionally, we will use outlier detection approaches, like Tukey's fences~\cite{tukey1977exploratory} and box plots, which do not require regression models to detect outliers. If there are outliers, we will analyse them by hand to gain insight into them. Finally, we may remove those data if we find enough evidence to do so after both outlier detection and manual evaluation.

\subsubsection{Measuring Effectiveness}
\label{subsec:measuringeffectiveness}
For rank-based tasks, i.e., \textit{task type B}, we will use Kendall's tau ranking distance formula~\cite{kendall1938new} to compute the distance between participant answers and the ground-truth. We report that number as the score between zero and one for those tasks. For list-based tasks, i.e., \textit{task types A and C}, like previous studies, we compute precision and recall~\cite{hattori2008precision}. As discussed, the goal of this study is to expose the change under different configuration settings and help developers to identify impacted deliverables for a specific configuration setting. To compute the correctness and completeness of the participant's Estimated Impacted Deliverables (EID), we compare them to Actual Impacted Deliverables (AID) using the following precision (correctness) and recall (completeness) formulas:
\[
Precision = \frac{EID \cap AID}{EID};
Recall = \frac{EID \cap AID}{AID}
\]
Due to the natural trade-off between precision and recall, we calculate the F-measure (i.e, the harmonic mean of precision and recall) to get an overall impression of task effectiveness.
%Precision measures the ratio of actual identified impacted deliverables to the number of total number of reported deliverables. However, recall measures the ratio of actual identified impacted deliverables to the total number of actual impacted deliverables.
\subsubsection{Descriptive Statistics} For each group, we will provide the mean, standard deviation, and relevant quantile values for our dependent variables and participant demographics. We will also include Spearman's $\rho$ pairwise correlation values to measure the strength of relations between the variables.

\subsubsection{Inferential Statistics} We will first use the Shapiro–Wilk test, along with a visual analysis, to determine if our data is normally distributed.

If our data follows non-normal distributions, we will use non-parametric statistical tests to answer RQ1 \& RQ2 because they impose fewer constraints on the distributions of analyzed data. In this case, we will use the Kruskal–Wallis test (i.e., One-way ANOVA on ranks) statistical hypothesis testing technique to identify whether there exists any statistically significant difference between the groups. If the test shows any difference among the groups, we will use the Dunn's Multiple Comparison Test~\cite{dunn1961multiple} to check for significant differences between the pairs while correcting the p-value with a Bonferroni adjustment. We will then apply the Cliff's Delta non-parametric effect size measure to assess the magnitude of the difference between the pairs.

If our data does follow a normal distribution, we will use a one-way ANOVA technique to identify whether a difference exists between the three groups. If a one-way ANOVA shows any differences, we will use Tukey's range test as a post-hoc test to explore differences between the \textit{No-Tool} group - \Tool group and \textit{Existing Tool} group - \Tool group while controlling the family-wise error. Finally, we will use Cohen's d for effect size calculations.

To reject the null hypothesis, the data should (1) present significant differences through the ANOVA test, (2) present significant differences through the post hoc test, and finally (3) show a large effect size between the \textit{No-Tool} group, the \textit{Existing tool} group, and the \Tool group. Due to the correction applied by the post hoc approaches, their results are conservative. Therefore, if they identify significant differences between the pairs, we can be confident of the results. On the other hand, if our data only shows significant differences through the ANOVA test, we will need to run the experiment with a larger sample size to make pairwise conclusions. If the number of recruited experienced developers is not large enough to perform the statistical analysis for that population, we will perform a preliminary analysis on it instead.
%~\maxNote{discuss post hoc technique mentioned by Shane: scottKnott}

%% file: variables_table.tex
\begin{table*}[ht]
\captionsetup{justification = centering}
\caption{The variables of the study}
\label{tbl:variables}
\begin{tabular}{>{\raggedright}p{3cm}| p{7.5cm}| p{1cm}|p{5cm}}
\hline
\textbf{Name}   & \textbf{Description}      & \textbf{Scale}   & \textbf{Operationalization}    \\           
\hline
\rowcolor{Gainsboro!60}
\multicolumn{4}{l}{\textit{Independent variables:}}  \\
Tooling level               & The tools available to the participants: no tool, existing tool, \Tool     & nominal       & See Section~\ref{subsec:variables}; randomized.\\
\hline
\rowcolor{Gainsboro!60}
\multicolumn{4}{l}{\textit{Dependent variables:}} \\

Number of correctly identified deliverables & Ratio of the   impacted deliverables correctly identified by the participants under a specific build-time configuration over the known impacted deliverables (RQ1) & ratio & Computed at the end using the harmonic mean (F-measure) for task types A \& C. See Sections~\ref{subsec:tasks} \&~\ref{subsec:measuringeffectiveness} \\
\rowcolor{Gainsboro!30}
Relative rate of correctly identified deliverables & Normalized pairwise disagreements between participant rankings of patches in terms of the number of impacted deliverables, and known correct rankings (RQ1)  & ratio & Calculated at the end for tasks of type B. See Section~\ref{subsec:measuringeffectiveness} 

\\
Exposure analysis effectiveness & The sum of the number of correctly identified deliverables and relative rate of correctly identified deliverables (RQ1) & ratio & Computed at the end using the number of correctly identified deliverables and the relative rate of correctly identified deliverables.
         \\
\rowcolor{Gainsboro!30}
Task time   &   The time needed for each participant to complete a task subtracting pauses (RQ2) & ratio & Measured by our web-based application. The participant can pause a task and resume manually. see Section~\ref{subsec:experimentui} 
\\
Exposure analysis efficiency & Ratio of the total score of the participant over the sum of all Task times (RQ2) & ratio & Total score is the sum of the scores of all of the individual tasks. see Section~\ref{subsec:measuringeffectiveness}                                                                             \\
\hline
\rowcolor{Gainsboro!60}
\multicolumn{4}{l}{\textit{Confounding/Measured variables:}}                                                                                                                                                                                                                           \\
CMake experience    &   Participant's experience in working with CMake build system     &   ordinal     &   Measured: 3-point scale (``none", ``tried", ``used in professional development"); questionnaire
\\
\rowcolor{Gainsboro!30}
Code changes                          & Changed code in diff format along with the other source files of the project                    & nominal & Design: each participant gets patches from three real-world projects  \\
\rowcolor{Gainsboro!30}
Configuration settings     & Environmental and build configuration settings of the build system: default configuration, custom                         & nominal       &  Design: for applicable tasks, each participant gets two configurations for build settings.   
\\
Current programming practice    &   How often the participant currently programs &   ordinal &   Measured: 3-point scale (``not", ``sometimes", ``often"); questionnaire
\\
\rowcolor{Gainsboro!30}
Development experience          & Participant's software development experience in years                                                                 & ordinal & Measured: 5-point scale (``less than a year" ... ``10 years or more"); questionnaire                                                           \\
Fitness &   Physical fitness of the participant, like tiredness, during the experiment  &   ordinal &   Measured: 5-point scale (``very tired" ... ``very fit"); questionnaire

\\
\rowcolor{Gainsboro!30}
Perceived task difficulty               & Participant's overall perception of the task provided during the experiment                                                   & ordinal   & Measured: 3-point scale (``easy", ``average", ``hard"); questionnaire at the end of each task              
\\
Project-specific experience     & Participant's past exprience with the provided project and patch & ordinal & Measured: 3-point scale (``none", ``user", ``contributor"); questionnaire at the end of each task

\end{tabular}
\end{table*}

%% file: threats.tex
\noindent\textit{Threats to internal validity:} Participants may vary in their capacity to estimate exposure. We strive to mitigate this by randomly assigning tasks to participants and by recruiting participants with varying levels of experience. Due to the challenges associated with obtaining a large sample of software developers, this study will focus on a statistically valid, but non-maximal number of participants. The pool of participants will retain enough statistical power to reject our null hypotheses. We are aware that the statistical power of the effect sizes of our findings is dependent on our final sample size and shall therefore endeavor to recruit as many participants as possible.

\noindent\textit{Threats to external validity:} We anticipate that most of our participants will volunteer from our three partner organizations. As such, they will likely have similar technological backgrounds. This might reduce the generalizability of our findings. To mitigate this effect, we also endeavor to obtain participants from other backgrounds.

\noindent\textit{Threats to construct validity:} Due to the Hawthorne effect, our participants are likely to behave differently in our experimental setting because they are aware that they are being monitored. We attempt to mitigate this threat by giving developers realistic tasks, letting them work on their own computers at a time and place of their choosing. Furthermore, we will not discuss the hypotheses of the study with the participants until after they completed their tasks. We are aware that our selected measurements do not fully capture the phenomena that we set out to measure (i.e., effectiveness and efficiency of assessing patch exposure). Nonetheless, we select a broad range of measurements and tasks that we believe to be meaningfully representative of the underlying phenomena of interest.